\documentstyle[amssymb,prl,aps,epsfig]{revtex}

\begin{document}

\title{Implications evinced by the phase diagram, anisotropy, magnetic
penetration depths, isotope effects and conductivities of cuprate
superconductors}
\author{T. Schneider and H. Keller}
\address{Physik-Institut der Universit\"{a}t Z\"{u}rich, Winterthurerstrasse 190,\\
CH-8057 Z\"{u}rich, Switzerland}
%\date{\today}
\maketitle

\begin{abstract}
Anisotropy, thermal and quantum fluctuations and their dependence
on dopant concentration appear to be present in all cuprate
superconductors, interwoven with the microscopic mechanisms
responsible for superconductivity. Here we review anisotropy,
in-plane and c-axis penetration depths, isotope effect and
conductivity measurements to reassess the universal behavior of
cuprates as revealed by the doping dependence of these phenomena
and of the transition temperature.
\end{abstract}

\bigskip

\begin{figure}[tbp]
\centering
\includegraphics[totalheight=6cm]{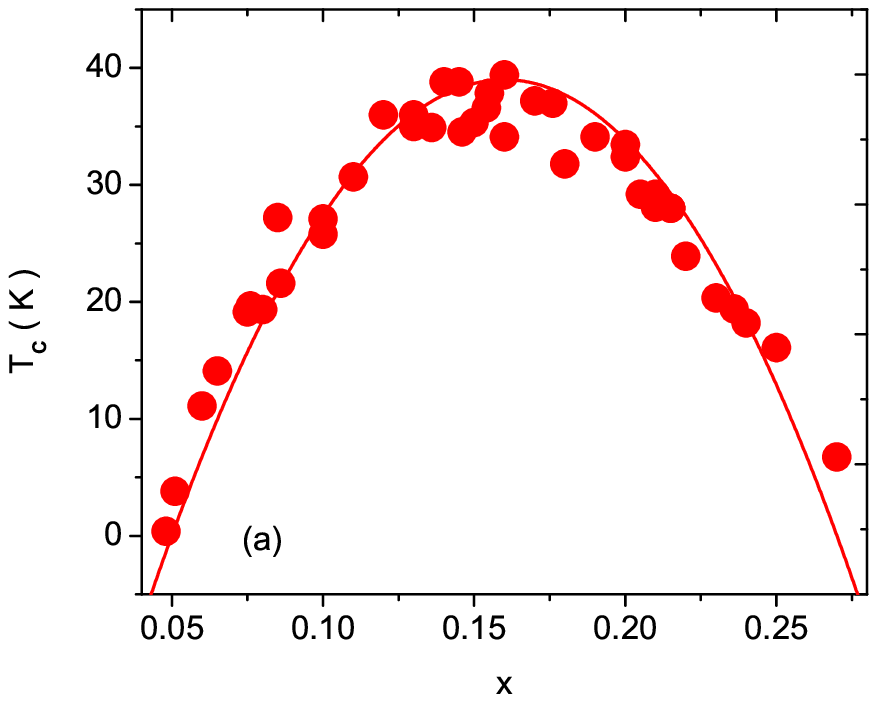}
\includegraphics[totalheight=6cm]{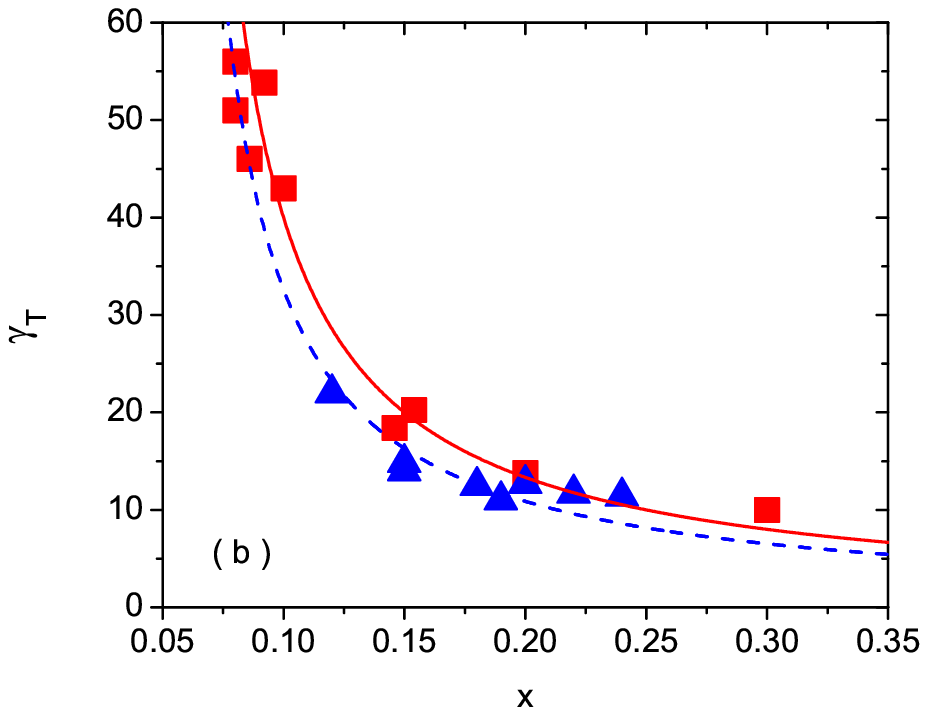}
\caption{(a) Variation of $T_{c}$ for La$_{2-x}$Sr$_{x}$CuO$_{4}$.
Experimental data taken from \protect\cite
{suzuki,nakamura,fukuzumi,willemin,kimura,sasagawa,hoferdis,shibauchi,panagopoulos}%
. The solid line is Eq.(\ref{eq1}) with $T_{c}\left( x_{m}\right)
=39$K. (b) $\gamma _{T}$ versus $x$ for
La$_{2-x}$Sr$_{x}$CuO$_{4}$. The squares are the experimental data
for $\gamma _{T_{c}}$ \protect\cite
{suzuki,nakamura,willemin,sasagawa,hoferdis} and the triangles for
$\gamma _{T=0}$ \protect\cite{shibauchi,panagopoulos}. The solid
curve and dashed lines are Eq.(\ref{eq2}) with $\gamma
_{T_{c},0}=2$ and $\gamma _{T=0,0}=1.63$.} \label{fig1}
\end{figure}

Establishing and understanding the phase diagram of cuprate
superconductors in the temperature-dopant concentration plane is
one of the major challenges in condensed matter physics.
Superconductivity is derived from the insulating and
antiferromagnetic parent compounds by partial substitution of ions
or by adding or removing oxygen. For instance La$_{2}$CuO$_{4}$
can be doped either by alkaline earth ions or oxygen to exhibit
superconductivity. The empirical phase diagram of
La$_{2-x}$Sr$_{x}$CuO$_{4}$ \cite
{suzuki,nakamura,fukuzumi,willemin,kimura,sasagawa,hoferdis,shibauchi,panagopoulos}
depicted in Fig. \ref{fig1}a shows that after passing the so
called underdoped limit $\left( x_{u}\approx 0.047\right) $,
$T_{c}$ reaches its maximum value $T_{c}\left( x_{m}\right) $ at
$x_{m}\approx 0.16$. With further increase of $x$, $T_{c}$
decreases and finally vanishes in the overdoped limit
$x_{o}\approx 0.273$. This phase transition line is thought to be
a generic property of cuprate superconductors \cite{tallon} and is
well described by the empirical relation
\begin{equation}
T_{c}\left( x\right) =T_{c}\left( x_{m}\right) \left( 1-2\left( \frac{x}{%
x_{m}}-1\right) ^{2}\right) =\frac{2T_{c}\left( x_{m}\right) }{x_{m}^{2}}%
\left( x-x_{u}\right) \left( x_{o}-x\right) ,  \label{eq1}
\end{equation}
proposed by Presland \emph{et al}.\cite{presland}. Approaching the
endpoints along the $x$-axis, La$_{2-x}$Sr$_{x}$CuO$_{4}$
undergoes at zero temperature doping tuned quantum phase
transitions. As their nature is concerned, resistivity
measurements reveal a quantum superconductor to insulator (QSI)
transition in the underdoped limit\cite
{momono,polen,book,klosters,tshk,tsphysB,parks} and in the
overdoped limit a quantum superconductor to normal state (QSN)
transition\cite{momono}.

\begin{figure}[tbp]
\centering
\includegraphics[totalheight=6cm]{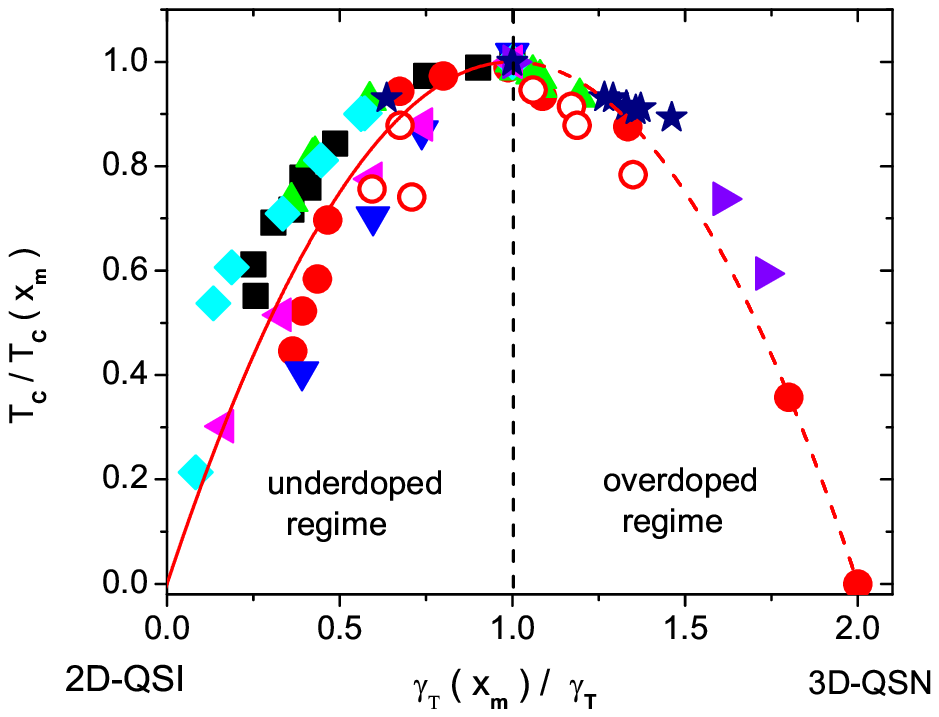}
\caption{$T_{c}\left( x\right) /T_{c}\left( x_{m}\right) $ versus
$\gamma _{T}\left( x_{m}\right) /$ $\gamma _{T}\left( x\right) $
for various cuprate families: La$_{2-x}$Sr$_{x}$CuO$_{4}$
($\bullet $, $T_{c}\left( x_{m}\right) =37$K, $\gamma
_{T_{c}}\left( x_{m}\right) =20$) \protect\cite
{suzuki,nakamura,willemin,sasagawa,hoferdis}\ , ($\bigcirc $,
$T_{c}\left( x_{m}\right) =37$K, $\gamma _{T=0}\left( x_{m}\right)
=14.9$) \protect\cite {shibauchi,panagopoulos},
HgBa$_{2}$CuO$_{4+\delta }$ ($\blacktriangle $, $T_{c}\left(
x_{m}\right) =95.6$K, $\gamma _{T_{c}}\left( x_{m}\right) =27$)
\protect\cite{hoferhg}, Bi$_{2}$Sr$_{2}$CaCu$_{2}$O$_{8+\delta }$
($\bigstar $, $T_{c}\left( x_{m}\right) =84.2$K, $\gamma
_{T_{c}}\left( x_{m}\right) =133$) \protect\cite{watauchi},
YBa$_{2}$Cu$_{3}$O$_{7-\delta }$ ($\blacklozenge $, $T_{c}\left(
x_{m}\right) =92.9$K, $\gamma _{T_{c}}\left( x_{m}\right) =8$)
\protect\cite{chien123},
YBa$_{2}$(Cu$_{1-y}$Fe$_{y}$)$_{3}$O$_{7-\delta }$ ($\blacksquare
$, $T_{c}\left( x_{m}\right) =92.5$K, $\gamma _{T_{c}}\left(
x_{m}\right) =9$)\protect\cite{chienfe},
Y$_{1-y}$Pr$_{y}$Ba$_{2}$Cu$_{3}$O$_{7-\delta }$
($\blacktriangledown $, $T_{c}\left( x_{m}\right) =91$K, $\gamma
_{T_{c}}\left( x_{m}\right) =9.3$)\protect\cite{chienpr},
BiSr$_{2}$Ca$_{1-y}$Pr$_{y}$Cu$_{2}$O$_{8}$ ($\blacktriangleleft
$, $T_{c}\left( x_{m}\right) =85.4$K, $\gamma _{T=0}\left(
x_{m}\right) =94.3$)\protect\cite{sun} and YBa$_{2}$(Cu$_{1-y}$
Zn$_{y}$)$_{3}$O$_{7-\delta }$ ($\blacktriangleright $,
$T_{c}\left( x_{m}\right) =92.5$K, $\gamma _{T=0}\left(
x_{m}\right) =9$)\protect\cite {panagopzn}. The solid and dashed
curves are Eq.(\ref{eq3}), marking the flow from the maximum
$T_{c}$ to QSI and QSN criticality, respectively.} \label{fig2}
\end{figure}

Another essential experimental fact is the doping dependence of
the anisotropy. In tetragonal cuprates it is defined as the ratio
$\gamma =\xi _{ab}/\xi _{c}$ of the correlation lengths parallel
$\left( \xi _{ab}\right)
$ and perpendicular $\left( \xi _{c}\right) $ to CuO$_{2}$ layers ($ab$%
-planes). In the superconducting state it can also be expressed as
the ratio $\gamma =\lambda _{c}/\lambda _{ab}$ of the London
penetration depths due to supercurrents flowing perpendicular
($\lambda _{c}$ ) and parallel ($\lambda _{ab}$ ) to the
$ab$-planes. Approaching a non-superconductor to superconductor
transition $\xi $ diverges, while in a superconductor to
non-superconductor transition $\lambda $ tends to infinity. In
both cases, however, $\gamma $ remains finite as long as the
system exhibits anisotropic but genuine 3D behavior. There are two
limiting cases: $\gamma =1$ characterizes isotropic 3D- and
$\gamma =\infty $ 2D-critical behavior. An instructive model where
$\gamma $ can be varied continuously is the anisotropic 2D Ising
model\cite{onsager}. When the coupling in the $y$ direction goes
to zero, $\gamma =\xi _{x}/\xi _{y}$ becomes infinite, the model
reduces to the 1D case, and $T_{c}$ vanishes. In the
Ginzburg-Landau description of layered superconductors the
anisotropy is related to the interlayer coupling. The weaker this
coupling is, the larger $\gamma $ is. The limit $\gamma =\infty $
is attained when the bulk superconductor corresponds to a stack of
independent slabs of thickness $d_{s}$. With respect to
experimental work, a considerable amount of data is available on
the chemical composition dependence of $\gamma $. At $T_{c}$ it
can be inferred from resistivity ($\gamma =\xi _{ab}/\xi
_{c}=\sqrt{\rho _{ab}/\rho _{c}}$) and magnetic torque
measurements, while in the superconducting state it follows from
magnetic torque and penetration depth ($\gamma =\lambda
_{c}/\lambda _{ab}$) data. In Fig. \ref{fig1}b we displayed the
doping dependence of $gamma _{T}$ evaluated at $T_{c}$ ($\gamma
_{T_{c}}$) and $T=0$ ($\gamma _{T=0}$). As the dopant
concentration is reduced, $\gamma _{T_{c}}$ and $\gamma _{T=0}$
increase systematically, and tend to diverge in the underdoped
limit. Thus the temperature range where superconductivity occurs
shrinks in the underdoped regime with increasing anisotropy. This
competition between anisotropy and superconductivity raises
serious doubts whether 2D mechanisms and models, corresponding to
the limit $\gamma _{T}=\infty $, can explain the essential
observations of superconductivity in the cuprates. From Fig.
\ref{fig1} it is also seen that $\gamma _{T}\left( x\right) $ is
well described by
\begin{equation}
\gamma _{T}\left( x\right) =\frac{\gamma _{T,0}}{x-x_{u}},
\label{eq2}
\end{equation}
where $\gamma _{T,0}$ is the quantum critical amplitude. Having
also other cuprate families in mind, it is convenient to express
the dopant concentration in terms of $T_{c}$. From Eqs.(\ref{eq1})
and(\ref{eq2}) we obtain the correlation between $T_{c}$ and
$\gamma _{T}$:
\begin{equation}
\frac{T_{c}\left( x\right) }{T_{c}\left( x_{m}\right) }=1-\left(
\frac{\gamma _{T}\left( x_{m}\right) }{\gamma _{T}\left( x\right)
}-1\right) ^{2},\ \ \gamma _{T}\left( x_{m}\right) =\frac{\gamma
_{T,0}}{x_{m}-x_{u}} \label{eq3}
\end{equation}
Provided that this empirical correlation is not merely an artefact
of La$_{2-x}$Sr$_{x}$CuO$_{4}$, it gives a universal perspective
on the interplay of anisotropy and superconductivity, among the
families of cuprates, characterized by $T_{c}\left( x_{m}\right) $
and $\gamma _{T}\left( x_{m}\right) $. For this reason it is
essential to explore its generic validity. In practice, however,
there are only a few additional compounds, including
HgBa$_{2}$CuO$_{4+\delta }$\cite{hoferhg}, for which the dopant
concentration can be varied continuously throughout the entire
doping range. It is well established, however, that the
substitution of magnetic and nonmagnetic impurities, depress
$T_{c}$ of cuprate superconductors very
effectively\cite{xiao,tarascon}. To compare the doping and
substitution driven variations of the anisotropy, we depicted in
Fig. \ref{fig2} the plot $T_{c}\left( x\right) /T_{c}\left(
x_{m}\right) $ versus $\gamma _{T}\left( x_{m}\right) /$ $\gamma
_{T}\left( x\right) $ for a variety of cuprate families. The
collapse of the data on the parabola, which is the empirical
relation (\ref{eq3}), reveals that this scaling form appears to be
universal. Thus, given a family of cuprate superconductors,
characterized by $T_{c}\left( x_{m}\right) $ and $\gamma
_{T}\left( x_{m}\right) $, it gives a universal perspective on the
interplay between anisotropy and superconductivity.

\begin{figure}[tbp]
\centering
\includegraphics[totalheight=6cm]{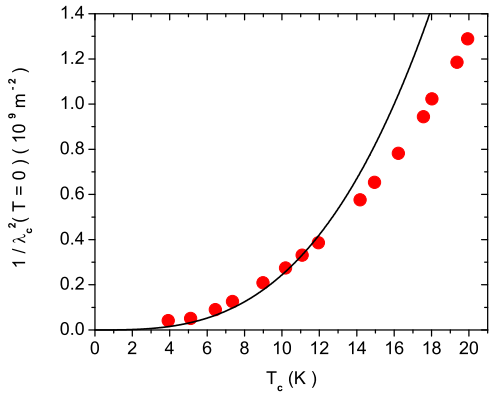}
\caption{$1/\lambda _{c}^{2}\left( T=0\right) $ \textit{vs.}
$T_{c}$ for YBa$_{2}$Cu$_{3}$O$_{6+x}$ single crystals, taken from
Hosseini \emph{et al}. \protect\cite {hosseini}. The solid line is
$1/\lambda _{c}^{2}\left( T=0\right) =2.44$ $10^{-4}T_{c}^{3}$ and
indicates the consistency with the 2D-QSI scaling relation
$T_{c}\propto \left( 1/\lambda _{c}^{2}\left( 0\right) \right)
^{z/(z+2)}$ and $z=1$.} \label{fig3}
\end{figure}

Close to 2D-QSI criticality various properties are not independent
but related by\cite{polen,book,klosters,tshk,tsphysB,parks}
\begin{equation}
T_{c}=\frac{\Phi _{0}^{2}R_{2}}{16\pi
^{3}k_{B}}\frac{d_{s}}{\lambda _{ab}^{2}\left( 0\right) }\propto
\gamma _{T}^{-z}\propto \delta ^{z\overline{\nu }},  \label{eq4}
\end{equation}
independently of the nature of the putative quantum critical
point. $k_{B}$ is the Boltzmann constant, $\Phi _{0}$ the
elementary flux quantum, $\lambda _{ab}\left( 0\right) $ the zero
temperature in-plane penetration depth, $z$ is the dynamic
critical exponent, $d_{s}$ the thickness of the sheets, and
$\overline{\nu }$ the correlation length critical exponent of the
2D-QSI transitions. $\delta $ measures the distance from the
critical point along the $x$ axis (see Fig.\ref{fig1}a), and
$R_{2}$ is a universal number. Since $T_{c}\propto d_{s}/\lambda
_{ab}^{2}\left( 0\right) \propto n_{s}^{\Box }$, where
$n_{s}^{\Box }$ is the aerial superfluid density, is a
characteristic 2D property, it also applies to the onset of
superfluidity in $^{4}$He films adsorbed on disordered substrates,
where it is well confirmed \cite{crowell}. A great deal of
experimental work has also been done in cuprates on the so called
Uemura plot, revealing an empirical correlation between $T_{c}$
and $d_{s}/\lambda _{ab}^{2}\left( 0\right) $\cite{uemura}.
Approaching 2D-QSI criticality, the data of a given family tends
to fall on a straight line, consistent with Eq.(\ref{eq4}).
Differences in the slope reflect the family dependent value of
$d_{s}$, the thickness of the sheets, becoming independent in the
2D limit\cite{book,klosters,tshk,tsphysB,parks}. The relevance of
$d_{s}$ was also confirmed in terms of the relationship between
the isotope effect on $T_{c}$ and $1/\lambda _{ab}^{2}$\cite
{tshk,tsiso}. Moreover, together with the scaling form (\ref{eq4})
the empirical relation (\ref{eq1}) implies 2D-QSI and 3D-QSN
transitions with $z\overline{\nu }=1$, while the empirical
behavior of the anisotropy (Eqs.(\ref {eq2}) and (\ref{eq3}))
require $\overline{\nu }=1$ at the 2D-QSI criticality. Thus, the
empirical correlations point to a 2D-QSI transition with $z=1$ and
$\overline{\nu }=1$.These estimates coincide with the theoretical
prediction for a 2D disordered bosonic system with long-range
Coulomb interactions, where $z=1$ and $\overline{\nu }\simeq
1$\cite {mpafisher,ca,herbutz1}. Here the loss of superfluidity is
due to the localization of the pairs, which ultimately drives the
transition. From the scaling relation (\ref{eq4}) it is seen that
measurements of the out of plane penetration depth of sufficiently
underdoped systems allow to estimate the dynamic critical exponent
$z$ directly, in terms of $T_{c}\propto \left( 1/\lambda
_{c}^{2}\left( 0\right) \right) ^{z/\left( z+2\right) }$, which
follows from Eq.(\ref{eq4}) with $\gamma _{T}=\lambda _{c}\left(
0\right) /\lambda _{ab}\left( 0\right) $. In Fig.\ref{fig3} we
displayed the data of Hosseini \emph{et al}.\cite{hosseini} for
heavily underdoped YBa$_{2}$Cu$_{3} $O$_{7-\delta }$ single
crystals. The solid line is $1/\lambda _{c}^{2}\left( T=0\right)
=2.44$ $10^{-4}T_{c}^{3}$ and uncovers the consistency with the
2D-QSI scaling relation $T_{c}\propto \left( 1/\lambda
_{c}^{2}\left( 0\right) \right) ^{z/\left( z+2\right) }$ with
$z=1$.

\begin{figure}[tbp]
\centering
\includegraphics[totalheight=6cm]{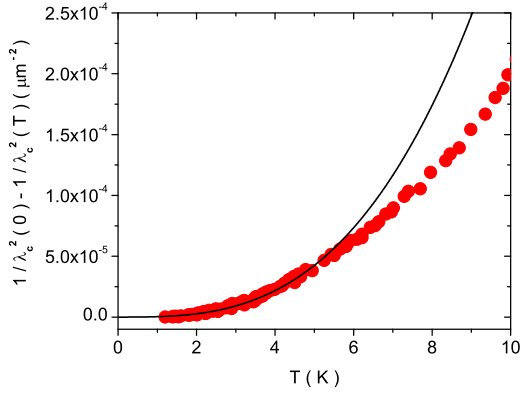}
\caption{$1/\lambda _{c}^{2}\left( T\right) -1/\lambda
_{c}^{2}\left( 0\right) $ \textit{vs.} $T$ for
YBa$_{2}$Cu$_{3}$O$_{6+x}$ single crystals, with $T_{c}$ values
$\approx $ $20.2$, $19.5$, $18.2$, $17.8$, $16.4$, and $15.1$K
taken from Hosseini \emph{et al}. \protect\cite{hosseini}. The
solid curve is $1/\lambda _{c}^{2}\left( T\right) -1/\lambda
_{c}^{2}\left( 0\right) =3.4$ $10^{-7}T^{3}$. } \label{fig4}
\end{figure}

Furthermore, Hosseini \emph{et al}.\cite{hosseini} found that at
low temperature $1/\lambda _{c}^{2}\left( T\right) -1/\lambda
_{c}^{2}\left( 0\right) $ is nearly doping independent. Given this
behavior close to the 2D-QSI transition, scaling predicts that in
leading order $1/\lambda _{c}^{2}\left( T\right) -1/\lambda
_{c}^{2}\left( 0\right) \propto T^{3}$ holds, in good agreement
with the experimental data displayed in Fig. \ref {fig4}. As the
in-plane penetration depth is concerned, there is mounting
experimental evidence that $1/\lambda _{ab}^{2}\left( T\right)
-1/\lambda _{ab}^{2}\left( 0\right) $ is in the limit
$T\rightarrow 0$ nearly doping independent as well\cite{klosters}.
In this case 2D-QSI scaling predicts in leading order $1/\lambda
_{ab}^{2}\left( T\right) -1/\lambda _{ab}^{2}\left( 0\right)
\propto T$, in agreement with the experimental data of numerous
cuprates\cite{klosters}.

We have seen that the doping tuned flow to the 2D-QSI critical
point is associated with a depression of $T_{c}$ and an
enhancement of $\gamma _{T}$. It implies that when the 2D-QSI
transition is approached, a no vanishing $T_{c}$ is inevitably
associated with an anisotropic but 3D condensation mechanism,
because $\gamma _{T}$ is finite for $T_{c}>0$ (see Figs. \ref
{fig1} and \ref{fig2}). This represents a serious problem for 2D
models\cite {anderson} as candidates to explain superconductivity
in the cuprates, and serves as a constraint on future work toward
a complete understanding. Note that the vast majority of
theoretical models focus on a single Cu-O plane, i.e., on the
limit of zero intracell and intercell c-axis coupling.

\begin{figure}[tbp]
\centering
\includegraphics[totalheight=6cm]{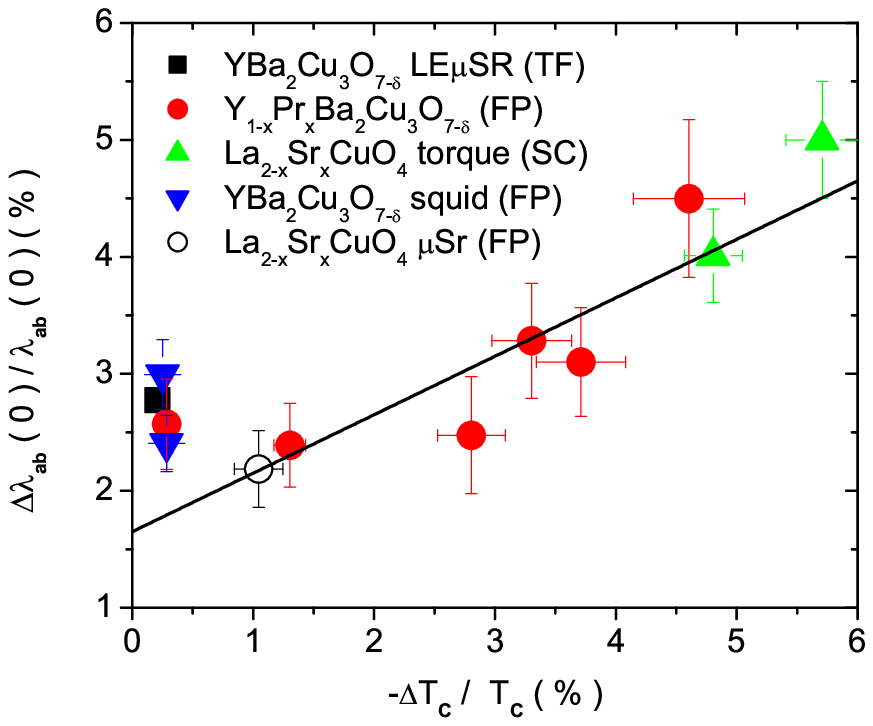}
\caption{Data for the oxygen isotope effect in underdoped
La$_{2-x}$Sr$_{x}$CuO$_{4}$($\bigcirc $:
x=0.15\protect\cite{zhao1}, $\blacktriangle $:x=0.08, 0.086
\protect\cite {hofer214},
Y$_{1-x}$Pr$_{x}$Ba$_{2}$Cu$_{3}$O$_{7-\delta }$ ($\bullet $: x=0,
0.2, 0.3, 0.4)\protect\cite{zhao1,khasanov123pr,rksite} and
YBa$_{2}$Cu$_{3}$O$_{7-\delta }$ ($\blacktriangledown $
\protect\cite{zhao1}, $\blacksquare $\protect\cite {khasanov123f})
\ in terms of $\Delta \left( \lambda _{ab}\left( 0\right) \right)
/$ $\lambda _{ab}\left( 0\right) $ \textit{vs.} -$\Delta
T_{c}/T_{c}$. The solid line indicates the flow to 2D-QSI
criticality and provides with Eq.(\ref{eq5}) an estimate for the
oxygen isotope effect on $d_{s}$, namely $\Delta
d_{s}/d_{s}=3.3(4)\%$.} \label{fig5}
\end{figure}

Since Eq.(\ref{eq4}) is universal, it also implies that the
changes $\Delta T_{c}$, $\Delta d_{s}$ and $\Delta \left(
1/\lambda _{ab}^{2}\left( T=0\right) \right) $, induced by
pressure or isotope exchange are not independent, but related by
\begin{equation}
\frac{\Delta T_{c}}{T_{c}}=\frac{\Delta d_{s}}{d_{s}}+\frac{\Delta
\left( 1/\lambda _{ab}^{2}\left( 0\right) \right) }{\left(
1/\lambda _{ab}^{2}\left( 0\right) \right) }=\frac{\Delta
d_{s}}{d_{s}}-2\frac{\Delta \left( \lambda _{ab}\left( 0\right)
\right) }{\lambda _{ab}\left( 0\right) }. \label{eq5}
\end{equation}
In particular, for the oxygen isotope effect ($^{16}$O vs.
$^{18}$O) of a physical quantity $X$ \ the relative isotope shift
is defined as $\Delta X/X=(^{18}X-^{16}X)/^{18}X$. In Fig.
\ref{fig5} we show the data for the oxygen isotope effect in
La$_{2-x}$Sr$_{x}$CuO$_{4}$\cite{hofer214,zhao1},
Y$_{1-x}$Pr$_{x}$Ba$_{2}$Cu$_{3}$O$_{7-\delta }$\cite
{zhao1,khasanov123pr,rksite} and YBa$_{2}$Cu$_{3}$O$_{7-\delta
}$\cite {zhao1,khasanov123f}, extending from the underdoped to the
optimally doped regime, in terms of $\Delta \left( \lambda
_{ab}\left( 0\right) \right) /$ $\lambda _{ab}\left( 0\right) $
\textit{vs.} $\Delta T_{c}/T_{c}$.
It is evident that there is a correlation between the isotope
effect on $T_{c}$ and $\lambda _{ab}\left( 0\right) $ which
appears to be universal for all cuprate families. Indeed, the
solid line indicates the flow to the 2D-QSI transition and
provides with Eq.(\ref{eq5}) an estimate for the oxygen isotope
effect on $d_{s}$, namely $\Delta d_{s}/d_{s}=3.3(4)\%$.
Approaching optimum doping, this contribution renders the isotope
effect on $T_{c}$ considerably smaller than that on $\lambda
_{ab}\left( 0\right) $. Indeed, even in nearly optimally doped
YBa$_{2}$Cu$_{3}$O$_{7-\delta }$, where $\Delta
T_{c}/T_{c}=-0.26(5)\%$, a substantial isotope effect on the
in-plane penetration depth, $\Delta \lambda _{ab}\left( 0\right)
/\lambda _{ab}\left( 0\right) =-2.8(1.0)\%$, has been established
by direct observation, using the novel low-energy muon-spin
rotation technique\cite{khasanov123f}. Note that these findings
have been obtained using various experimental techniques on
powders, thin films and single crystals.

In this context it is important to recognize that the substantial
isotope effect on the in-plane penetration depth is incompatible
with the Migdal-Eliashberg (ME)\cite{me} of the electron-phonon
interaction\cite {varenna}. Since the lattice parameters remain
essentially unaffected\cite {conder,raffa} by isotope exchange,
while the dynamics associated with the mass of the respective ions
is modified, the substantial isotope effect on the zero
temperature penetration depth requires a renormalization of the
normal state Fermi velocity \textbf{v}$_{F}\ \rightarrow \ \widetilde{%
\mathbf{v}}_{F}=\ $\textbf{v}$_{F}/(1+f)$ where \textbf{v}$_{F}$
is the bare velocity and $f$ the electron-phonon coupling
constant. However, in the Migdal-Eliashberg \cite{me} treatment of
the electron-phonon interaction $f$ is independent of the ionic
masses and assumed to be small \cite{maksimov,carbotte}. This is
true if the parameter $\omega _{0}f/E_{F}\ $is small, where
$\omega _{0}$ is the relevant phonon frequency and $E_{F}$ the
Fermi energy. Thus the isotope effect on the penetration depth is
expected to be small, of the order of the adiabatic parameter
$\widetilde{\gamma }=\omega _{0}/E_{F}<<1$. The ME theory retains
terms only of order 0. Cuprates, however, have Fermi energies much
smaller than those of conventional metals \cite{randeria} so that
$ \widetilde{\gamma }$ is no longer negligible small. In any case
the substantial isotope effect on the in-plane penetration depth
uncovers the coupling between local lattice distortions and
superfluidity and the failure of the Migdal-Eliashberg (ME)
treatment of the electron-phonon interaction, predicting
$1/\lambda ^{2}\left( 0\right) $ to be independent of the ionic
masses\cite{me}. Evidence for this coupling emerges from the
oxygen isotope effect on $d_{s}$, the thickness of the
superconducting sheets, while the lattice parameters remain
unaffected. Indeed, the relative shift, $\Delta d_{s}/d_{s}\approx
3.3(4)\%$, apparent in Fig.\ref{fig5}, implies local oxygen
distortions which do not modify the lattice parameters but are
coupled to the superfluid. Although the majority opinion on the
mechanism of superconductivity in the cuprates is that it occurs
via a purely electronic mechanism involving spin excitations, and
lattice degrees of freedom are supposed to be irrelevant, the
relative isotope shift $\Delta d_{s}/d_{s}\approx 3.3(4)\%$
uncovers clearly the existence and relevance of the coupling
between the superfluid and local lattice distortions.

Finally we turn to the finite temperature critical behavior. Close
to the critical temperature $T_{c}$ of the superconductor to
normal state transition, in a regime roughly given by the Ginzburg
criterion\cite{book,ffh,tsda,tshkws}, order parameter fluctuations
dominate critical properties. In recent years, the effect of the
charge of the superconducting order parameter in this regime in
three dimensions has been studied extensively\cite
{coleman,halperin,dasgupta,folk,kleinert,herbut,herbut2,olsson,mo}.
While for strong type-I materials, the coupling of the order
parameter to transverse gauge field fluctuations is expected to
render the transition first order \cite{halperin}, it is
well-established that strong type-II materials should exhibit a
continuous phase transition, and that sufficiently close to
$T_{c}$, the charge of the order parameter is relevant
\cite{folk,kleinert,herbut,herbut2,olsson,hove,mo}. However, in
cuprate superconductors within the fluctuation dominated regime,
the region close to $T_{c}$, where the system crosses over to the
regime of charged fluctuations turns out to be too narrow to
access. For instance, optimally doped
YBa$_{2}$Cu$_{3}$O$_{7-\delta }$, while possessing an extended
regime of critical fluctuations, is too strongly type-II to
observe charged critical
fluctuations\cite{book,ffh,tsda,tshkws,kamal}. Indeed, the
effective dimensionless charge $\widetilde{e}=\xi /\lambda
=1/\kappa $ is small in strongly type II superconductors ($\kappa
>>1$). The crossover upon approaching $T_{c}$ is thus initially to
the critical regime of a weakly charged superfluid where the
fluctuations of the order parameter are essentially those of an
uncharged superfluid or XY-model \cite{ffh}. Furthermore, there is
the inhomogeneity induced finite size effect which renders the
asymptotic critical regime unattainable\cite{tsrkhk,tsbled,tsdc}.
However, underdoped cuprates could open a window onto this new
regime because $\kappa _{ab}$ is expected to become rather small.
As outlined above, in this regime cuprates undergo a quantum
superconductor to insulator
transition\cite{book,klosters,tshk,tsphysB,parks} and correspond
to a 2D disordered bosonic system with long-range coulomb
interactions. Close to this quantum transition $T_{c}$, $\lambda
_{ab}$ and $\xi _{ab}$ scale as $T_{c}\propto \lambda
_{ab}^{-2}\propto \xi _{ab}^{-z}\propto \lambda
_{c}^{-2(z+2)/z}$\cite{book,klosters,tshk,tsphysB,parks}, while
$\xi _{c}\rightarrow d_{s}$. These relations yield with the
dynamic critical exponent
$z=1$\cite{book,parks,mpafisher,ca,herbutz1}, $\kappa _{ab}\propto
T_{c}^{1/2}$ and $\kappa _{c}\propto T_{c}^{-3/2}$. Noting that
$T_{c}$ decreases by approaching the underdoped limit, the
in-plane penetration depth appears to be a potential candidate to
observe charged criticality in sufficiently homogeneous and
underdoped cuprates, while the c-axis counterpart is expected to
exhibit neutral critical behavior.

\begin{figure}[tbp]
\centering
\includegraphics[totalheight=6cm]{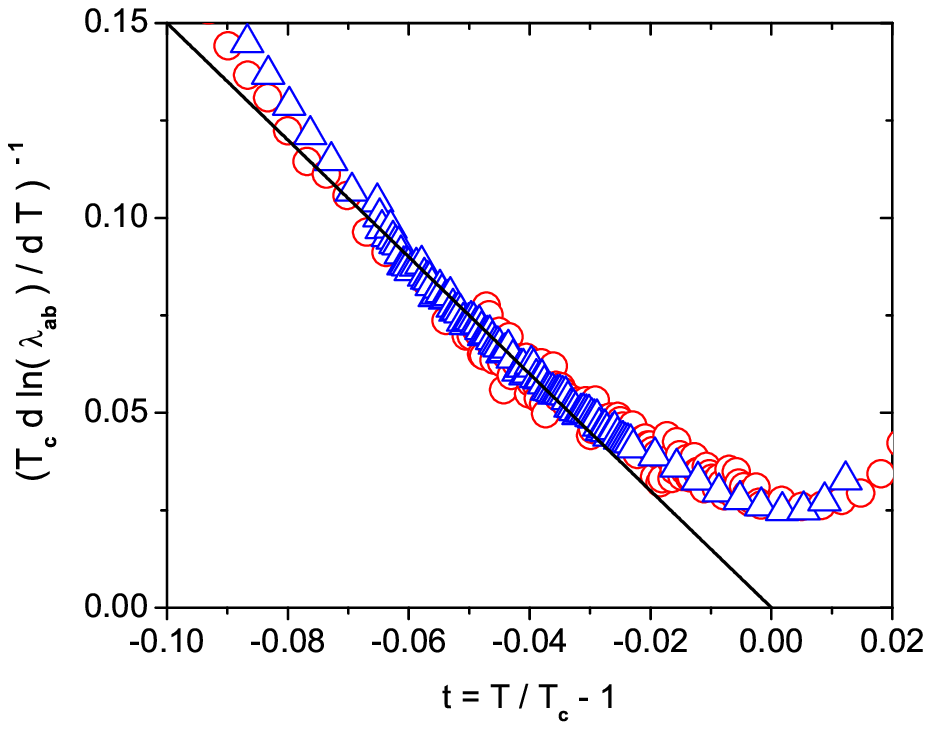}
\caption{$(T_{c}d\ln \lambda _{ab}/dT)^{-1}$ with $\lambda _{ab}$
in $\mu $m \textit{vs.} $t$ for YBa$_{2}$Cu$_{3}$O$_{6.59}$ under
hydrostatic pressure of $0.5 $kbar\ $\left( \triangle \right) $and
$10.5$kbar $\left( \bigcirc \right) $ taken from
ref.\protect\cite{tsrk2}. The straight line with slope $1/\nu
\simeq 3/2$ corresponds according to Eq. (\ref{eq6}) to charged
critical behavior with $T_{c}=57.82$K and $T_{c}=61.1$K.}
\label{fig6}
\end{figure}

When charged fluctuations dominate the in-plane penetration depth
and the correlation length are related
by\cite{herbut,herbut2,olsson,hove,mo}
\begin{equation}
\lambda _{ab}=\kappa _{ab}\xi _{ab},\text{ }\lambda _{ab}=\lambda
_{0ab}\left| t\right| ^{-\nu },\text{ }\nu \simeq 2/3,
\label{eq6}
\end{equation}
contrary to the uncharged case, where $\lambda \propto \sqrt{\xi
}$ and with that
\begin{equation}
\text{ }\lambda _{ab}=\lambda _{0ab}\left| t\right| ^{-\nu /2},
\label{eq7}
\end{equation}
where $t=T/T_{c}-1$. In a plot $(d\ln \lambda _{ab}/dT)^{-1}$
\textit{vs.} $T$ charged critical behavior is then uncovered in
terms of a temperature range where the data falls on a line with
slope $1/\nu \simeq 3/2$, while in the neutral (3D-XY) case it
collapses on a line with slope $2/\nu \simeq 3$. Clearly, in an
inhomogeneous system the phase transition is rounded and $(d\ln
\lambda _{ab}/dT)^{-1}$ does not vanish at $T_{c}$. In Fig.
\ref{fig6} we displayed $(T_{c}d\ln \lambda _{ab}/dT)^{-1}$
\textit{vs.} $t$ for YBa$_{2}$Cu$_{3}$O$_{6.59}$ with applied
hydrostatic pressures of $0.5$kbar\ and $10.5$kbar, derived from
the measured in-plane penetration depth data of ref.\cite {tsrk2}.
The solid line with slope $1/\nu \simeq 3/2$ indicates according
to Eq. (\ref{eq6}) the charged criticality in homogeneous systems
with $T_{c}=57.82$K and $T_{c}=61.1$K. Although the charged
criticality is attained there is no sharp transition because
$\lambda _{ab}$ does not diverge at $T_{c}$. Inhomogeneities
prevent the correlation length $\xi _{ab}=\lambda _{ab}/\kappa
_{ab}$ to grow beyond the spatial extent $L_{ab}$ of the
homogenous domains in the $ab$-plane. For a discussion of the
associated finite size effect we refer to ref.\cite{tsrk2}. In any
case these measurements clearly reveal that the temperature
dependence of the in-plane penetration depth opens a window onto
charged criticality at finite temperature in underdoped cuprates.

Extended c-axis penetration depth measurements on
YBa$_{2}$Cu$_{3}$O$_{6+x}$ single crystals with $T_{c}$'s ranging
from $4$ to $20$K have been performed by Hosseini \emph{et
al.}\cite{hosseini}. In Fig. \ref{fig6} we displayed their data
for the sample with $T_{c}\approx 19.75$K in terms $\lambda
_{c}^{-2}$and $(d\ln \lambda _{c}^{-2}/dT)^{-1}$ \textit{vs.} $T$.
The solid curve is $\lambda _{c}^{-2}=\lambda _{0c}^{-2}\left(
1-T/Tc\right) ^{\nu }$ with $T_{c}=19.75$K and $\lambda
_{0c}^{-2}=2.15$ (10$^{9}$m$^{-2}$) and the dashed one $(d\ln
\lambda _{c}^{-2}/dT)^{-1}=-1/\nu \left( T_{c}-T\right) $ with
$\nu \simeq 2/3$.
\begin{figure}[tbp]
\centering
\includegraphics[totalheight=6cm]{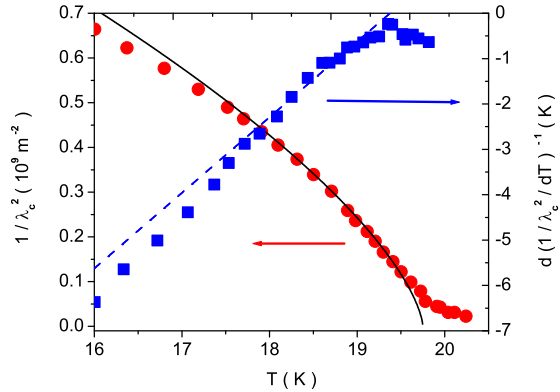}
\caption{$\lambda _{c}^{-2}$and $(d\ln \lambda _{c}^{-2}/dT)^{-1}$
\textit{vs.} $T$ for a YBa$_{2}$Cu$_{3}$O$_{6+x}$ single crystal
derived from the data of Hosseini \emph{et
al.}\protect\cite{hosseini}. The solid curve is  $\lambda
_{c}^{-2}=\lambda _{0c}^{-2}\left( 1-T/Tc\right) ^{\nu }$ with
$T_{c}=19.75$K and  $\lambda _{0c}^{-2}=2.15$ (10$^{9}$m$^{-2}$)
and the dashed one $(d\ln \lambda _{c}^{-2}/dT)^{-1}=-1/\nu \left(
T_{c}-T\right) $ with $\nu \simeq 2/3$. They indicate according to
Eq. (\ref{eq7}) the leading uncharged (3D-XY) critical behavior.}
\label{fig7}
\end{figure}
They indicate according to Eq. (\ref{eq7}) the leading uncharged
(3D-XY) critical behavior. Noting that in this underdoped regime
the anisotropy $\gamma =\lambda _{c}/\lambda _{ab}$ is rather
large (see Fig. \ref{fig2}) the critical regime where thermal 3D
fluctuations dominate will be compared to optimum doping much
narrower. Here it extends to $t\approx 77/93.7-1\simeq
-0.18$\cite{kamal}. Nevertheless, a glance to
Fig. \ref{fig7} shows that the 3D-XY critical regime is attained for $18$K$%
\lesssim T\lesssim 19$K, corresponding to $0.09\gtrsim \left|
t\right| \gtrsim 0.04$. The tail above $19$K is attributable to an
inhomogeneity
induced finite size effect. Indeed, according to the scaling relation $%
\kappa _{c}\propto T_{c}^{-3/2}$ it is unlikely that the crossover
to charged criticality is attainable. In contrast and consistent
with $\kappa _{ab}\propto T_{c}^{1/2}$ charged criticality turned
out to be accessible in the $ab$-plane penetration depth.

\begin{figure}[tbp]
\centering
\includegraphics[totalheight=6cm]{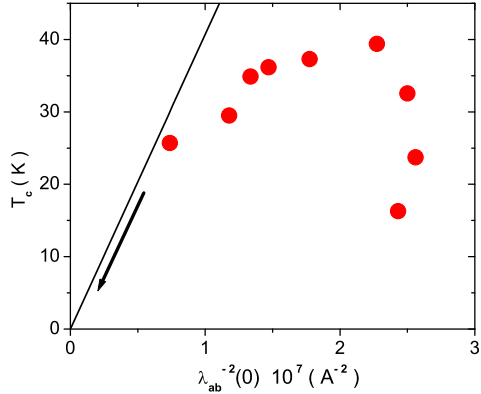}
\caption{$T_{c}$ \textit{vs.} $1/\lambda _{ab}^{2}\left( 0\right)
$ for La$_{2-x}$Sr$_{x}$CuO$_{4}$. Data taken from Uemura \emph{et
al.} \protect\cite {uemura,uemura214} and Panagopoulos \emph{et
al.} \protect\cite{panagopoulos}. The straight line is
Eq.(\ref{eq4}) with $R_{2}d_{s}=6.5$\AA\ and the arrow indicates
the flow to 2D-QSI transition criticality.} \label{fig8}
\end{figure}

We have seen that the linear relationship between $T_{c}$ and
$1/\lambda _{ab}^{2}\left( 0\right) $ in the underdoped regime,
referred to as the Uemura relation, is a consequence of the the
dimensional crossover associated with the flow to the 2D-QSI
transition (see Eq.(\ref{eq4}). To illustrate this behavior we
displayed in Fig. \ref{fig8} $T_{c}$ \textit{vs.} $1/\lambda
_{ab}^{2}\left( 0\right) $ for La$_{2-x}$Sr$_{x}$CuO$_{4}$. The
straight line is Eq.(\ref{eq4}) in terms of $T_{c}=25\lambda
_{ab}^{-2}\left( 0\right) $ and the arrow indicates the flow to
the 2D-QSI criticality. However, approaching the optimally doped
regime, where $T_{c}$ reaches its maximum value (see Fig.
\ref{fig1}), the observed $T_{c}$'s are systematically lower than
the 2D-QSI line and this trend continues in the overdoped regime.

\begin{figure}[tbp]
\centering
\includegraphics[totalheight=6cm]{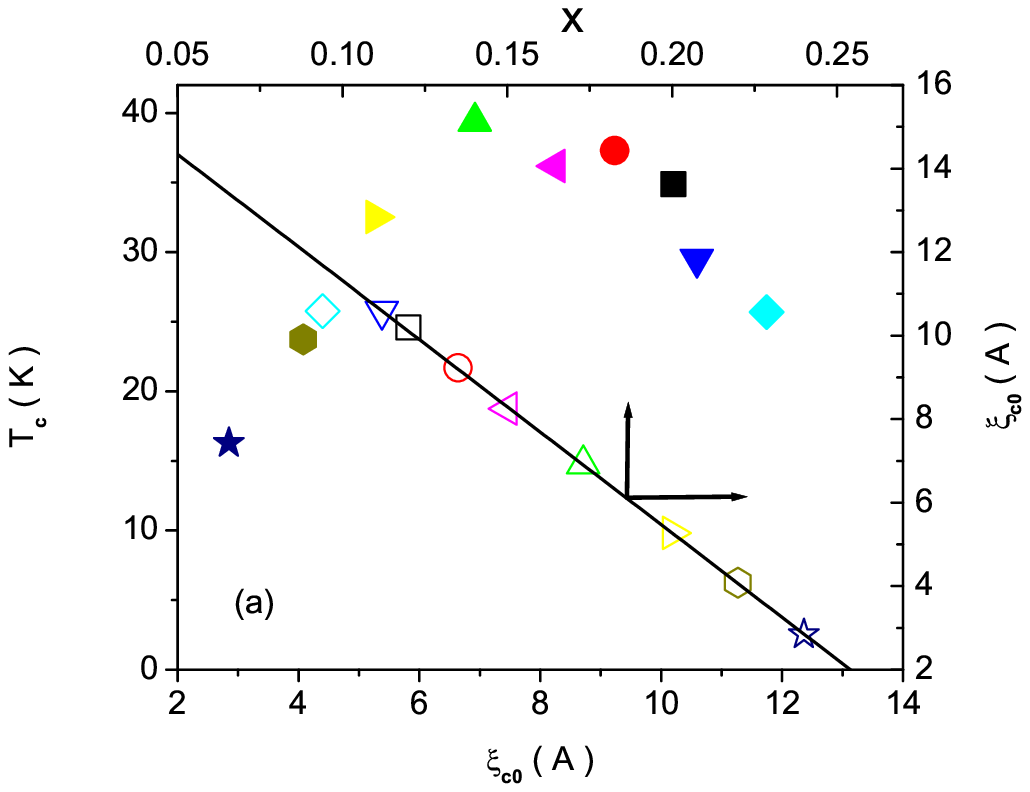}
\includegraphics[totalheight=6cm]{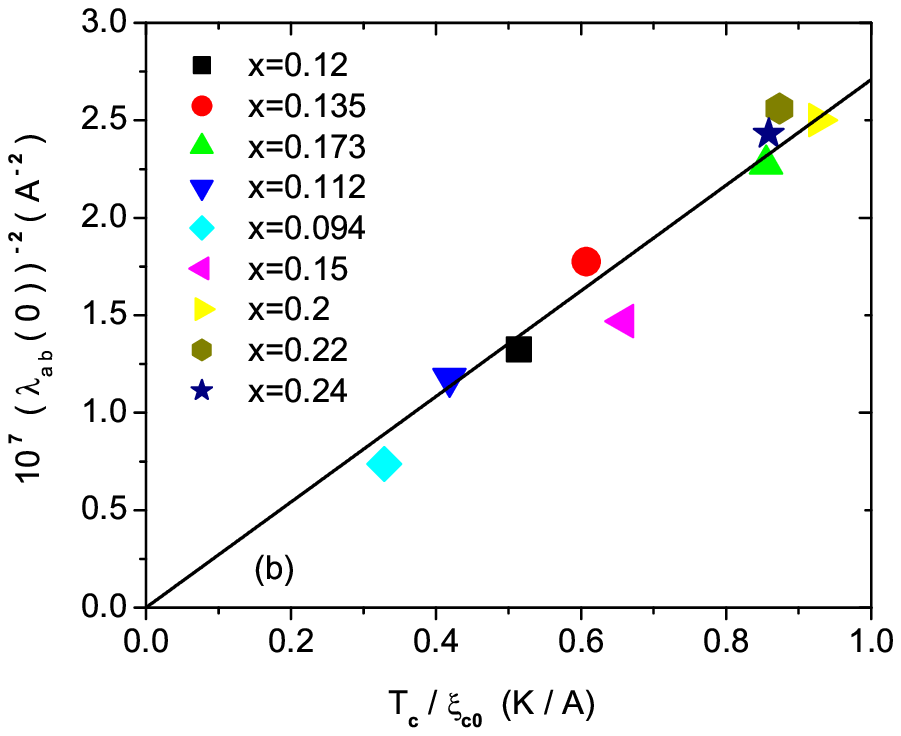}
\caption{(a) $T_{c}\ $\textit{vs.} $\xi _{c0}$ and $\xi _{c0}$
\textit{vs. }$x$ for La$_{2-x}$Sr$_{x}$CuO$_{4}$. Data taken from
Pangopoulos \emph{et al.} \protect\cite{panagopoulos} and
Shibauchi \emph{et al.} \protect\cite{shibauchi}. The solid line
is Eq.(\ref{eq11c}). (b) $1/\lambda _{ab}^{2}\left( 0\right) \
$\textit{vs.} $T_{c}/\xi _{c0}$ for the same data with $\xi _{c0}$
given by Eq.(\ref {eq11c}). The straight line is  $1/\lambda
_{ab}^{2}\left( 0\right) =2.71$ $T_{c}/\xi _{c0}$.} \label{fig9}
\end{figure}

To provide an understanding we invoke 3D-XY universality expected
to hold along the phase transition line, $T_{c}\left( x\right) $
in La$_{2-x}$Sr$_{x} $CuO$_{4}$ (see Fig. \ref{fig1}), as long as
the charge of the pairs can be neglected. In this case the
transition temperature $T_{c}$ and the critical amplitudes of the
penetration depths $\lambda _{ab0}$ and transverse correlation
lengths $\xi _{ab0}^{tr}$ are related by\cite{book,parks}
\begin{equation}
k_{B}T_{c}=\frac{\Phi _{0}^{2}}{16\pi ^{3}}\frac{\xi
_{ab0}^{tr}}{\lambda _{ab0}^{2}}=\frac{\Phi _{0}^{2}}{16\pi
^{3}}\frac{\xi _{c0}^{tr}}{\lambda _{c0}^{2}},  \label{eq8}
\end{equation}
where $\lambda _{i}^{2}\left( T\right) =\lambda _{i0}^{2}\left(
1-T/T_{c}\right) ^{-\nu }$ and $\xi _{i}^{t}\left( T\right) =\xi
_{i0}^{t}\left( 1-T/T_{c}\right) ^{-\nu }$ with $\nu \simeq 2/3$.
For our purpose it is convenient to express the transverse
correlation length to the correlation lengths above $T_{c}$ in
terms of\cite{book,parks,peliasetto}
\begin{equation}
\frac{\xi _{ab0}^{tr}}{\xi _{c0}}=f\approx 0.453,\text{ }\frac{\xi _{c0}^{tr}%
}{\xi _{ab0}}=\gamma f,  \label{eq9}
\end{equation}
where the anisotropy is given by
\begin{equation}
\gamma ^{2}=\left( \frac{\lambda _{c0}}{\lambda _{ab0}}\right) ^{2}=\frac{%
\xi _{c0}^{tr}}{\xi _{ab0}^{tr}}=\left( \frac{\xi _{ab0}}{\xi
_{co}}\right) ^{2}.  \label{eq10}
\end{equation}
Combining Eqs.(\ref{eq8}) and (\ref{eq9}) we obtain the universal
relation
\begin{equation}
T_{c}\lambda _{ab0}^{2}=\frac{\Phi _{0}^{2}f}{16\pi ^{3}k_{B}}\xi
_{c0}. \label{eq11}
\end{equation}
It holds,  as long as cuprates fall into the 3D-XY universality
class,
irrespective of the doping dependence of $T_{c}$, $\lambda _{ab0}^{2}$ and $%
\xi _{c0}$. Fo this reason it provides a sound basis for universal
plots. However, there is the serious drawback that reliable
experimental estimates for $T_{c}$ and the critical amplitudes
$\lambda _{ab0}$ and $\xi _{c0}$ measured on the same sample are
not yet available. Nevertheless, some progress can be made by
noting that in the underdoped regime, approaching
the 2D-QSI transition the universal scaling forms (\ref{eq4}) and (\ref{eq11}%
) should match. This requires
\begin{equation}
f\xi _{c0}\rightarrow Rd_{s},  \label{eq11a}
\end{equation}
so that away from 2D-QSI criticality
\begin{equation}
T_{c}\lambda _{ab}^{2}\left( 0\right) \simeq \frac{\Phi
_{0}^{2}f}{16\pi ^{3}k_{B}}\xi _{c0},  \label{eq11b}
\end{equation}
holds. Thus, when both $T_{c}$ and $1/\lambda _{ab}^{2}\left(
0\right) $ increase, $T_{c}$ values below $T_{c}=\left( \Phi
_{0}^{2}R_{2}/6\pi ^{3}k_{B}\right) d_{s}/\lambda _{ab}^{2}\left(
0\right) $ (Eq.(\ref{eq4})) require $\xi _{c0}$ to fall off from
its limiting value $\xi _{c0}=\left( R_{2}/f\right) d_{s}$. The
doping dependence of $\xi _{c0}$ in La$_{2-x}$Sr$_{x}$CuO$_{4}$,
deduced from Eq.(\ref{eq11b}) and the experimental data for
$T_{c}$ and $\lambda _{ab}^{2}\left( 0\right) $ is displayed in
Fig.\ref{fig9}a in terms of $T_{c}\ $\textit{vs.} $\xi _{c0}$ and
$\xi _{c0}$ \textit{vs. }$x$ . Approaching the underdoped limit
$\left( x\approx 0.05\right) $, where $T_{c}$ vanishes
(Fig.\ref{fig1}) and the 2D-QSI transition occurs, $\xi _{c0}$
increases nearly linearly with decreasing $x$ to approach a fixed
value. Indeed, the data is consistent with
\begin{equation}
\xi _{c0}=\left( 16\pi ^{3}k_{B}/\left( \Phi _{0}^{2}f\right)
\right) T_{c}\lambda _{ab}^{2}\left( 0\right) \approx
14.34-60.47(x-0.05)\text{\AA }, \label{eq11c}
\end{equation}
yielding the limiting value $\xi _{c0}\left( x=0.05\right) \approx
14.34 \text{\AA\ and }f\xi _{c0}\left( x=0.05\right)
=R_{2}d_{s}\approx 6.5$\AA\ used in Fig. \ref{fig8}. An essential
result is that $\xi _{c0}$ adopts in the underdoped limit
($x\simeq 0.05$) where the 2D-QSI transition occurs and $T_{c}$
vanishes its maximum value $\xi _{c0}\approx 14.34$\AA , which is
close to the c-axis lattice constant $c\simeq 13.29$\AA . Thus, a
finite transition temperature requires a reduction of $\xi _{c0}$,
well described over an unexpectedly large doping range by
Eq.(\ref{eq11c}). Noting that this relation transforms with
Eq.(\ref{eq2}) to
\begin{equation}
\xi _{c0}\approx 14.34-60.47\gamma _{0}/\gamma \text{\AA },
\label{eq11d}
\end{equation}
this behavior is intimately connected to the doping dependence of
the anisotropy. Hence, a finite $T_{c}$ requires unavoidably a
finite anisotropy $\gamma $. The lesson is, in agreement with the
empirical relation (\ref{eq3}), that superconductivity in
La$_{2-x}$Sr$_{x}$CuO$_{4}$ is an anisotropic but 3D phenomenon
which disappears in the 2D limit. From the plot $1/\lambda
_{ab}^{2}\left( 0\right) \ $\textit{vs.} $T_{c}/\xi _{c}$
displayed in Fig. \ref{fig9}b, where according to Eq.(\ref{eq11b})
universal behavior is expected to occur, the data is seen to fall
rather well on a straight line. This is significant, as moderately
underdoped, optimally and overdoped La$_{2-x}$Sr$_{x}$CuO$_{4}$
falling according to Fig. \ref{fig8} well off the 2D-QSI behavior
$T_{c}\propto 1/\lambda _{ab}^{2}\left( 0\right) $ now scale
nearly onto a single line. Thus the approximate 3D-XY scaling
relation (\ref {eq11b}), together with the empirical doping
dependence of the c-axis correlation length (Eqs.(\ref{eq11c}) and
(\ref{eq11d})) are consistent with the available experimental data
for La$_{2-x}$Sr$_{x}$CuO$_{4}$ and uncovers the relevance of the
anisotropy. However, the linear doping dependence of $\xi _{c}$ is
not expected to hold closer to the overdoped limit $\left(
x\approx 0.27\right) $ where $T_{c}$ vanishes and a 3D quantum
superconductor to normal state (3D-QSN) transition is expected to
occur (see Fig.\ref{fig1}).

Since Eq.(\ref{eq11}) is universal, it should hold for all
cuprates falling in the accessible critical regime into the 3D-XY
universality class, irrespective of the doping dependence of
$T_{c}$, $\lambda _{ab0}^{2}$, $\xi _{ab0}$, $\gamma $ and $\xi
_{c0}$. Having seen that charged criticality is accessible in the
heavily underdoped regime only, Eq.\ref{eq11}) rewritten in the
form
\begin{equation}
\frac{\xi _{c0}}{\lambda _{ab0}^{2}T_{c}}=\frac{\xi _{ab0}\gamma
}{\lambda _{c0}^{2}T_{c}}=\frac{16\pi ^{3}k_{B}}{\Phi _{0}^{2}f},
\label{eq12}
\end{equation}
provides a sound basis for universal plots. However, as
aforementioned there is the serious drawback that reliable
experimental estimates for the the critical amplitudes and the
anisotropy at $T_{c}$ measured on the same sample and for a
variety of cuprates are not yet available. Nevertheless, as in the
case of La$_{2-x}$Sr$_{x}$CuO$_{4}$ outlined above, progress can
be made by invoking the approximate scaling form (\ref{eq11b}) and
by expressing the correlation lengths in terms of $\sigma
_{i}^{dc}$ the real part of the frequency dependent conductivity
$\sigma _{i}^{dc}\left( \omega \right) $ in direction $i$
extrapolated to zero frequency at $T\gtrsim T_{c}$\cite{homes123}
in terms of
\begin{equation}
\text{ }1/\xi _{c}\simeq \sigma _{ab}^{dc}/s_{ab},\text{ }1/\left(
\gamma \xi _{ab}\right) \simeq \sigma _{c}^{dc}/s_{c},
\label{eq13}
\end{equation}
$s_{i}$ with dimension $\Omega ^{-1}$incorporates the temperature
dependence. With that the universal relation (\ref{eq12})
transforms with Eq.(\ref{eq11b}) to
\begin{equation}
\frac{1}{\lambda _{ab}^{2}\left( 0\right) T_{c}\sigma
_{ab}^{dc}}\simeq
\frac{1}{\lambda _{c}^{2}\left( 0\right) T_{c}\sigma _{c}^{dc}}\simeq \frac{%
16\pi ^{3}k_{B}}{\Phi _{0}^{2}fs_{ab}},  \label{eq14}
\end{equation}
because $\sigma _{ab}^{dc}/\sigma _{c}^{dc}=\gamma ^{2}$ and with that $%
s_{ab}=s_{c}$. To check this relation we note that the 2D-QSI scaling from (%
\ref{eq4}) transforms with the relation $d_{s}=\sigma
^{sheet}/\sigma
_{ab}^{dc}$ between sheet conductivity $\sigma ^{sheet}$ and conductivity $%
\sigma _{ab}^{dc}$ to
\begin{equation}
\frac{1}{\lambda _{ab}^{2}\left( 0\right) T_{c}\sigma _{ab}^{dc}}=\frac{%
16\pi ^{3}k_{B}}{\Phi _{0}^{2}R_{2}\sigma ^{sheet}},\text{ \
}\sigma
^{sheet}=\frac{h}{4e^{2}}\sigma _{0}\simeq \sigma _{0}1.55\text{ }%
10^{-4}\Omega ^{-1},  \label{eq15}
\end{equation}
where $h/4e^{2}=6.45$k$\Omega $ is the quantum of resistance and
$\sigma _{0} $ is a dimensionless constant of order
unity\cite{herbutsig}. Thus,
\begin{equation}
\frac{1}{\lambda _{ab}^{2}\left( 0\right) }\simeq \frac{10.3\text{ }10^{-5}}{%
R_{2}\sigma _{0}}T_{c}\sigma _{ab}^{dc},  \label{eq16}
\end{equation}
with  $\lambda _{ab}\left( 0\right) $ in $\mu $m, $T$ in K and
$\sigma
_{ab}^{dc}$ in ($\Omega ^{-1}$cm$^{-1}$), and the structure of the $ab$%
-expression in Eq.(\ref{eq14}) is recovered. Similarly,
approaching 2D-QSI criticality $\lambda _{c}\left( 0\right) $ and
$\sigma _{c}^{dc}$ scale as \cite{tseuro}
\begin{equation}
\lambda _{c}\left( 0\right) =\Omega _{s}\left( \sigma
_{c}^{dc}\right) ^{-\left( 2+z\right) /4},  \label{eq17}
\end{equation}
where $z=1$ is the dynamic critical exponent of the 2D-QSI transition. $%
\Omega _{s}$ is a non-universal factor of proportionality. Noting
that in this limit $\sigma _{c}^{dc}$ and $T_{c}$ scale as
$T_{c}\propto \left( \sigma _{c}^{dc}\right) ^{z/2}$ the
$c$-expression in Eq.(\ref{eq14}) is readily recovered.
Accordingly, there is no contradiction between the scaling forms
(\ref{eq14}) and (\ref{eq17}). In this context it should be kept
in mind that the universal relation is valid for the neutral case
only.
However we have seen that the effective dimensionless charge $\widetilde{e}%
_{ab}=1/\kappa _{ab}$ scales as $\widetilde{e}_{ab}\propto
T_{c}^{-1/2}$ and the charged critical regime becomes accessible.
This is not the case for the $c$-axis penetration depth because
$\widetilde{e}_{c}\propto T_{c}^{3/2}$. For this reason, as the
underdoped limit (2D-QSI transition) is approached
there should be a crossover from Eq.(\ref{eq14}) to the universal relation (%
\ref{eq15}), becoming manifest in different constants on the right
hand side.

On the other hand, approaching the 3D quantum superconductor to
normal state (QSN) transition $\lambda _{ab,c}\left( 0\right) $,
$T_{c}$ and $\sigma _{ab,c}^{dc}$ scale as\cite{parks} $1/\lambda
_{ab,c}^{2}\left( 0\right) \propto T_{c}^{\left( 1+z\right) /z}$,
$\sigma _{ab}^{dc}\propto T_{c}^{-\left( z_{cl}-1\right) /z}$,
where $z$ is the dynamic critical exponent of this quantum
transition and $z_{cl}$ the exponent associated with the finite
temperature critical dynamics. Indeed, in this limit the
correlation length $\xi _{\tau }$ associated with the finite
temperature critical dynamics cannot be eliminated. Since $\xi
_{\tau }$ scales as $\xi _{\tau }\propto \xi _{ab}^{z_{cl}}$ we
obtain $\sigma _{c}^{dc}\propto \xi _{c}\xi _{\tau }/\xi
_{ab}^{2}\propto \xi _{\tau }/\left( \gamma \xi _{ab}\right)
\propto \xi _{ab}^{z_{cl}-1}\propto T_{c}^{-\left( z_{cl}-1\right)
/z}$. Accordingly,
\begin{equation}
\lambda _{ab}^{2}\left( 0\right) T_{c}\sigma _{ab}^{dc}\propto
\lambda _{c}^{2}\left( 0\right) T_{c}\sigma _{c}^{dc}\propto
T_{c}^{-z_{cl}/z} \label{eq18}
\end{equation}
should hold close to 3D-QSN criticality. Furthermore in this
regime the effective charge becomes negligible small so that 3D-XY
scaling applies.
Indeed $\widetilde{e}_{ab,c}=1/\kappa _{ab,c}$ scales as $\widetilde{e}%
_{ab,c}\propto T_{c}^{\left( z-1\right) /2z}$ because $\lambda
_{ab,c}\propto T_{c}^{-\left( z+1\right) /2z}$ and $\xi
_{ab,c}\propto T_{c}^{-1/z}$. A potential candidate for the 3D-QSN
transition is the model proposed by Herbut\cite{herbutz2}. It
describes the disordered d-wave superconductor to disordered metal
transition at weak coupling and is characterized by the dynamic
critical exponents $z=2$. When this holds true the effective
charge scales at $\widetilde{e}_{ab,c}\propto T_{c}^{-3/4}$ and
3D-XY scaling is no longer applicable. Furthermore there is
experimental
evidence for $z_{cl}=2$\cite{tshkws,osborn}. Accordingly, the scaling form (%
\ref{eq14}) does not hold in the overdoped regime close to 3D-QSN
criticality. In this regime Eq.(\ref{eq17}) is replaced
by\cite{tseuro}
\begin{equation}
\lambda _{c}\left( 0\right) \propto \left( \sigma _{c}^{dc}\right) ^{\frac{%
1+z}{2\left( z_{cl}-1\right) }},  \label{eq19}
\end{equation}
with a non-universal factor of proportionality. What is
particularly remarkable is then that the 3D-XY scaling from
(\ref{eq14}) predicts that
all points of $1/\lambda _{ab}^{2}\left( 0\right) $ \textit{vs}. $%
T_{c}\sigma _{ab}^{dc}$ and $1/\lambda _{c}^{2}\left( 0\right) $
\textit{vs.} $T_{c}\sigma _{c}^{dc}$ should fall onto a single
line with the exception of the overdoped regime where the scaling
form (\ref{eq18}) applies.

\begin{figure}[tbp]
\centering
\includegraphics[totalheight=6cm]{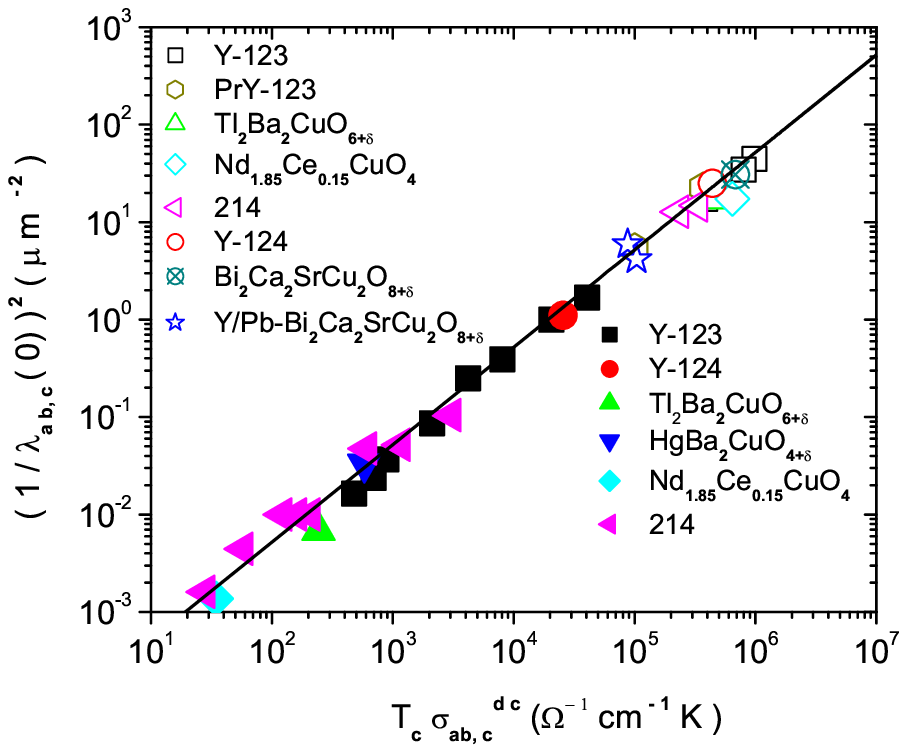}
\caption{$1/\lambda _{ab}^{2}\left( 0\right) $ \textit{vs.}
$T_{c}\sigma _{ab}^{dc}$ (open symbols) and $1/\lambda
_{c}^{2}\left( 0\right) $ \textit{vs.} $T_{c}\sigma _{c}^{dc}$
(full symbols) for various cuprates as collected by Homes \emph{et
al.}\protect\cite{homesuni}. The straight line is $1/\lambda
_{ab,c}^{2}\left( 0\right) =5.2$ $10^{-5}$ $T_{c}\sigma
_{ab,c}^{dc}$. The experimental data is taken from
\protect\cite{10,11,12,19,20,21} for YBa$_{2}$Cu$_{3}$O$_{7-\delta
}$ (Y-123), \protect\cite{12} for
Pr/Pb-YBa$_{2}$Cu$_{3}$O$_{7-\delta }$(Pr/Pb-123),
\protect\cite{19,13} for YBa$_{2}$Cu$_{4}$O$_{8}$ (Y-124),
\protect\cite{19,13} for Tl$_{2}$Ba$_{2}$CuO$_{6+\delta }$,
\protect\cite{12,22} for Bi$_{2}$Ca$_{2}$SrCu$_{2}$O$_{8+\delta }$
and Y/Pb Bi$_{2}$Ca$_{2}$SrCu$_{2}$O$_{8+\delta }$,
\protect\cite{5,6} for Nd$_{1.85}$Ce$_{0.15}$CuO$_{4}$,
\protect\cite {homesuni,14} for La$_{2-x}$Sr$_{x}$CuO$_{4}$(214),
and \protect\cite{homesuni} for HgBa$_{2}$CuO$_{4+\delta }$.}
\label{fig10}
\end{figure}

In Fig.\ref{fig10} we displayed $1/\lambda _{ab}^{2}\left(
0\right) $ \textit{vs}. $T_{c}\sigma _{ab}^{dc}$ and $1/\lambda
_{c}^{2}\left( 0\right) $ \textit{vs.} $T_{c}\sigma _{c}^{dc}$ as
collected by by Homes \emph{et al.} \cite{homesuni}. In agreement
with the scaling from (\ref{eq14}) the $ab$-plane and $c$-axis
data appears to be well described by the same line, namely
$1/\lambda _{ab,c}^{2}\left( 0\right) \simeq 5.2\
10^{-5}T_{c}\sigma _{ab,c}^{dc}$, yielding for $R_{2}\sigma _{0}$
in the 2D-QSI relation (Eq.(\ref{eq16})) the estimate $R_{2}\sigma
_{0}\cong 1.98$. Furthermore, all points of $1/\lambda
_{ab}^{2}\left( 0\right) $ \textit{vs.} $T_{c}\sigma _{ab}^{dc}$
(open symbols) nearly fall onto a single line with a slope of
unity. This is significant, as moderately underdoped, optimally
and overdoped materials, which fell well off of the 2D-QSI
behavior $T_{c}\propto 1/\lambda _{ab}^{2}\left( 0\right) $ (see
Eq.(\ref{eq4}) and Fig.\ref{fig8}) now scale nearly onto a single
line, in agreement with Fig. \ref{fig9}b.

This agreement demonstrates that the approximate scaling relation
\ref{eq14} captures the essentials of the exact 3D-XY scaling form
(\ref{eq12}) by eliminating the doping dependence of the critical
amplitudes towards the 2D-QSI transition in terms of there zero
temperature counterparts. This evidence for anisotropic 3D-XY
scaling raises again serious doubts that 2D models \cite{anderson}
are potential candidates to explain superconductivity in the
cuprates. However the data does not extend sufficiently close to
2D-QSI- and 3D-QSN criticality in order to uncover the
aforementioned crossovers from the scaling from (\ref{eq14}) to
(\ref{eq15}) and from (\ref{eq14}) to (\ref{eq18}).

\begin{figure}[tbp] \centering
\includegraphics[totalheight=6cm]{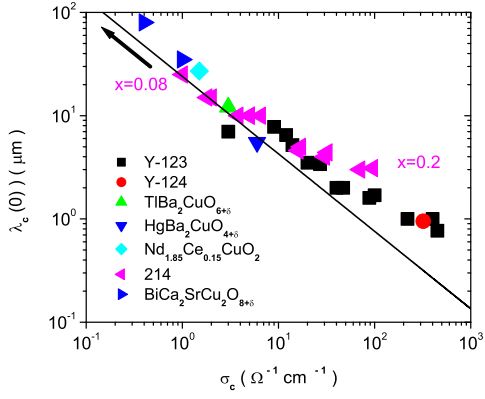}
\caption{$1/\lambda _{c}\left( 0\right) $ \textit{vs.} $\sigma
_{c}^{dc}$ using the data shown in Fig. \ref{fig10} and taken from
\protect\cite{uchida,motohashi} for YBa$_{2}$Cu$_{3}$O$_{7-\delta
}$ (Y-123), La$_{2-x}$Sr$_{x}$CuO$_{4}$(214), and from
\protect\cite{shibata} for Bi$_{2}$Ca$_{2}$SrCu$_{2}$O$_{8+\delta
}$. The straight line corresponds to Eq.(\ref{eq17}) \ with
$\Omega _{s}=24$ appropriate for
La$_{2-x}$Sr$_{x}$CuO$_{4}$(214).} \label{fig11}
\end{figure}

In contrast, the plot $\lambda _{c}\left( 0\right) $ \textit{vs.
}$\sigma _{c}^{dc}$ displayed in Fig. \ref{fig11} provides
according to the scaling form (\ref{eq17}) information on the flow
to 2D-QSI criticality. The straight line is Eq.(\ref{eq17}) with
$\Omega _{s}=24$ appropriate for La$_{2-x}$Sr$_{x}$CuO$_{4}$(214)
and the arrow indicate the direction of this flow. The data for
La$_{2-x}$Sr$_{x}$CuO$_{4}$ covers the range from $x=0.08$ to
$0.2$. Although the data is sparse one anticipates that the
deviations from the straight line behavior increase systematically
as the overdoped region is approached. As noted previously
\cite{tseuro} this behavior is attributable to the initial
crossover to 3D-QSN-criticality, where the scaling form
(\ref{eq19}) applies. Clearly more experimental data extending
much closer to the overdoped limit are needed to uncover this
crossover. Actually this also applies to the crossover from the
scaling form (\ref{eq14}) to (\ref{eq18}).

Since Eq.(\ref{eq14}), with the exception of the heavily overdoped
regime, turned out to be nearly universal, it also implies that
the changes $\Delta T_{c}$, $\Delta \left( 1/\lambda
_{ab,c}^{2}\right) $ and $\Delta \sigma _{ab,c}^{dc}$, induced by
pressure or isotope exchange are not independent, but related by
\begin{equation}
\frac{\Delta \left( 1/\lambda _{ab,c}^{2}\left( 0\right) \right)
}{\left( 1/\lambda _{ab,c}^{2}\left( 0\right) \right)
}=\frac{\Delta T_{c}}{T_{c}}+\frac{\Delta \sigma
_{ab,c}^{dc}}{\sigma _{ab,c}^{dc}}. \label{eq20}
\end{equation}
In particular, for the oxygen isotope effect ($^{16}$O
\textit{vs.} $^{18}$O) of a physical quantity $X$ \ the relative
isotope shift is defined as $\Delta X/X=(^{18}X-^{16}X)/^{18}X$.
Since close to 2D-QSI criticality $d_{s}=\sigma ^{sheet}/\sigma
_{ab}^{dc}$ and with that $\Delta \sigma _{ab}^{dc}/\sigma
_{ab}^{dc}=-\Delta d_{s}/d_{s}$ we recover the universal relation
(\ref{eq5}). Close to optimum doping where in
YBa$_{2}$Cu$_{3}$O$_{7-\delta }$ $\Delta T_{c}/T_{c}=-0.26(5)\%$
and $\Delta \left( 1/\lambda _{ab,c}^{2}\left( 0\right) \right)
/\left( 1/\lambda _{ab,c}^{2}\left( 0\right) \right)
=-5.6(2.0)\%$\cite{khasanov123f} it predicts $-5.6\%$ effect on
$\Delta \sigma _{ab}^{dc}/\sigma _{ab}^{dc}$ evaluated above and
extrapolated to $T_{c}$. Hence, the absence of a substantial
isotope effect on the transition temperature of cuprates does not
rule out phonons as the bosons responsible for the coupling
between the charge carriers, as previously suggexted\cite{franck}.

To summarize we observed remarkable agreement between the
experimental data of underdoped cuprates and the flow to the
2D-QSI critical endpoint in the underdoped limit. Although an
inhomogeneity induced finite size effect makes the asymptotic
critical regime unattainable, there is considerable evidence that
the 2D-QSI transition falls into the universality class of a 2D
disordered bosonic system with long-range Coulomb interactions.
The loss of superfluidity is due to the localization of the pairs,
which ultimately drives the transition. Important implications
include: (i) A finite transition temperature and superfluid
density in the ground state are unalterably linked to a finite
anisotropy. This finding raises serious doubts that 2D models
\cite{anderson} are potential candidates to explain
superconductivity in cuprates.(ii) The doping dependence of \ the
large oxygen isotope effects on the zero temperature in-plane
penetration depth confirms the flow to 2D-QSI criticality, poses a
fundamental challenge to this understanding and calls for a theory
that goes beyond Migdal-Eliashberg. Although the majority opinion
on the mechanism of superconductivity in the cuprates is that it
occurs via a purely electronic mechanism involving spin
excitations, and lattice degrees of freedom are supposed to be
irrelevant, the relative isotope shift of the thickness of the
superconducting sheets $\Delta d_{s}/d_{s}\approx 3.3(4)\%$
uncovers clearly the existence and relevance of the coupling
between the superfluid and local lattice distortions.(iii) Given
the striking feature that at low temperature the absolute change
in $1/\lambda _{ab}^{2}$ and $1/\lambda _{c}^{2}$ has essentially
no doping dependence, in contrast to $1/\lambda _{ab}^{2}\left(
0\right) $ and $1/\lambda _{c}^{2}\left( 0\right) $, scaling
around the 2D-QSI critical endpoint implies a $T^{3}$ power law
for $1/\lambda _{c}^{2}\left( T\right) $ and $T$ for $1/\lambda
_{ab}^{2}\left( T\right) $, consistent with experiment. Such power
law behavior also follows from the nodes characteristic of a
d-wave energy gap in the one particle density of states. However,
our derivation shows that in underdoped cuprates these power laws
are characteristic features of the 2D-QSI transition at the
critical endpoint and hold for both s- and d-wave pairing . (iv)
Similarly, we have seen that the large changes of $1/\lambda
_{ab}^{2}\left( 0\right) $, $1/\lambda _{c}^{2}\left( 0\right) $
and $T_{c}$, consistent with the relation $T_{c}\propto \lambda
_{ab}^{-2}\left( T=0\right) \propto \lambda _{ab}^{-2/3}\left(
T=0\right) $, follow from the flow to the 2D-QSI criticality as
well. (v) Because the Ginzburg parameters scale as $\kappa
_{ab}\propto T_{c}^{1/2}$ and $\kappa _{c}\propto T_{c}^{-3/2}$
this flow was shown to open a door to observe charged criticality
in the in-plane penetration depth at finite temperature. An
intriguing implications for microscopic models is then the
increasing value of the effective charge of the pairs in the
ab-plane, where $\widetilde{e}_{ab}=1/\kappa _{ab}\propto
T_{c}^{-1/2}$, while in the c-direction it disappears as
$\widetilde{e}_{c}=1/\kappa _{c}\propto T_{c}^{3/2}$. vi) We
demonstrated here that the empirical relation $\lambda
_{ab}^{2}\left( 0\right) T_{c}\sigma _{ab}^{dc}=\lambda
_{c}^{2}\left( 0\right) T_{c}\sigma _{c}^{dc}\simeq const$
\cite{homesuni} is a consequence of 3D-XY universality extended to
anisotropic systems. Nevertheless, these relations clearly reveal
that the absence of a substantial isotope effect on the transition
temperature of cuprates does not ruled out phonons as the bosons
responsible for coupling between the charge carriers, as
previously suggested.

\acknowledgments The authors are grateful to D. Di Castro, R.
Khasanov, S. Kohout, K.A. M\"{u}ller, J. Roos and A. Shengelaya
for very useful comments and suggestions on the subject matter.
This work was partially supported by the Swiss National Science
Foundation and the NCCR program \textit{Materials with Novel
Electronic Properties} (MaNEP) sponsored by the Swiss National
Science Foundation.

\end{document}